\documentclass[sigplan]{acmart}
\acmConference[PriSC'22]{6th Workshop on Principles of Secure Compilation}{January 22, 2022}{Philadelphia, PA, USA}
\acmYear{2022}
\acmISBN{}
\acmDOI{}

\setcopyright{none}

\usepackage{prooftree}

\newcommand{\deffont}[1]{\textbf{#1}}
\newcommand{\fall}[1]{\forall{#1}.}
\newcommand{\lam}[1]{\lambda{#1}.}

\title{The Supervisionary proof-checking kernel}
\subtitle{Or: a work-in-progress toward proof-generating code}

\author{Dominic P. Mulligan}
\orcid{}
\affiliation{
  \institution{Systems Research Group, Arm Research}
  \streetaddress{Fulbourn Road}
  \city{Cambridge}
  \country{United Kingdom}
}
\email{dominic.mulligan@arm.com}

\author{Nick Spinale}
\orcid{}
\affiliation{
  \institution{Systems Research Group, Arm Research}
  \streetaddress{Fulbourn Road}
  \city{Cambridge}
  \country{United Kingdom}
}
\email{nick.spinale@arm.com}

\begin{document}

\maketitle

\paragraph{Some scene setting}

Interactive theorem proving software is typically designed around a trusted proof-checking \deffont{kernel}, the sole system component capable of authenticating theorems.
Untrusted automation procedures reside outside of the kernel, and drives it to deduce new theorems via an API.
Kernel and untrusted automation are typically implemented in the same programming language---the ``meta-language''---usually some functional programming language in the ML family.
This strategy---introduced by Milner in his LCF proof assistant~\cite{Milner1972LogicFC}---is a reliability mechanism, aiming to ensure that any purported theorem produced by the system is indeed entailed by the theory within the logic.

Changing tack, operating systems are also typically designed around a trusted kernel, a privileged component responsible for---amongst other things---mediating interaction betwixt user-space software and hardware.
Untrusted processes interact with the system by issuing kernel \deffont{system calls} across a hardware privilege boundary.
In this way, the operating system kernel \deffont{supervises} user-space processes.

Though ostensibly very different, squinting, we see that the two kinds of kernel are tasked with solving the same task: enforcing system invariants in the face of interaction with untrusted code.
Yet, the two solutions to solving this problem, employed by the respective kinds of kernel, are very different.
In this abstract, we explore designing proof-checking kernels as \deffont{supervisory software}, where separation between kernel and untrusted code is enforced by \emph{privilege}.

\paragraph{System interface}

\emph{Supervisionary} is a proof-checking system for Gordon's HOL, structured as supervisory software, and provides a \deffont{system interface} to untrusted code similar to an operating system's system call interface.
Supervisionary is implemented as a WebAssembly~\cite{DBLP:conf/pldi/HaasRSTHGWZB17} (Wasm henceforth) host, allowing us to prototype rapidly.

We use Rust as our implementation language, rather than a functional programming language.
This entails no risk of unsoundness providing our system interface is carefully designed.
In particular, the kernel manages various private heaps within which \deffont{kernel objects} are allocated, corresponding to the paraphenalia of any HOL implementation---type-formers, types, term constants, terms, and theorems---and never directly exposed to untrusted code.
Kernel objects are allocated in response to system calls like:
\begin{displaymath}
\mathtt{Term.Handle.AllocateApplication(left, right, out)}
\end{displaymath}
Here, both $\mathtt{left}$ and $\mathtt{right}$ are \deffont{kernel handles}, assumed to point-to allocated terms, whilst $\mathtt{out}$ points-to a buffer in untrusted code's memory.
If neither $\mathtt{left}$ nor $\mathtt{right}$ dangle, and the types of their referents match, a fresh handle is generated which points-to a new HOL term application object, with internal pointers to the functional- and argument-terms.
This handle is returned to the caller via the $\mathtt{out}$ pointer.

The manipulation and querying of kernel objects is performed defensively by the kernel itself on behalf of untrusted code.
The kernel is careful to maintain invariants such as the inductivity of its heaps, with nodes in the kernel object graph pointing-to allocated objects at all times.

Space constraints prevent us from describing the entire Supervisionary system interface for working with, and on, kernel objects.
However, note that theorems are also constructed in a similar way to terms, inductively building derivation trees.
For example, the HOL symmetry rule is exposed as:
\begin{displaymath}
\mathtt{Theorem.Handle.AllocateSym(pre, out)}
\end{displaymath}
Here, $\mathtt{pre}$ points-to an existing theorem object $\Gamma \vdash r = s$ and after succeeding, passing obvious checks, $\mathtt{out}$ contains a handle that points-to a newly-allocated theorem $\Gamma \vdash s = r$.

Note that one interesting consequence of this style of implementation is the ability to provide concise specifications of Supervisionary's system interface functions.
Essentially, the Supervisionary kernel is a grand exercise in pointer manipulation, and as such our system call specifications can be expressed as Hoare Triples, using Separation Logic~\cite{DBLP:conf/lics/Reynolds02} as our assertion language.
Writing $\mathtt{h} \mapsto_{\mathtt{trm}} \mathtt{Application(l, r)}$ to assert that the handle $\mathtt{h}$ points-to a term application (of the term pointed-to by $l$ to the term pointed-to by $r$), and writing $\mathtt{out} \mapsto \mathtt{b}$ to assert that $\mathtt{out}$ points-to the Boolean value $\mathtt{b}$, we have:
\begin{gather*}
\{ h \mapsto_{\mathtt{trm}} \mathtt{Application(l, r)} \} \\
\mathtt{Term.Handle.IsApplication(h, out)} \\
\{ \mathtt{out} \mapsto \mathtt{True} \}
\end{gather*}
(Here, the triple $\{ P \} C \{ Q \}$ asserts that if the command $C$ executes in a state concordant with $P$ then the command succeeds and produces a state concordant with $Q$.)

Note that Supervisionary lacks any analogue of the traditional LCF meta-language.
Kernel and automation are now decoupled, and code written in \emph{any} language can ``drive'' the kernel, providing it produces binary-compatible executables.

\vspace{\baselineskip}
We now turn to speculation around potential uses of Supervisionary.
Exploring what follows is still a work-in-progress.
\paragraph{Runtime verification}

Given our use of Wasm, we could extend our system interface by also implementing the Wasm System Interface~\cite{Wasi}---a POSIX-like interface for Wasm.
This would transform Supervisionary from a mere programmable proof-checker into a general-purpose sandbox, capable of executing arbitrary programs, albeit with an unusual interface for constructing proofs.

\vspace{\baselineskip}
But: \emph{what happens if we blur the lines between Supervisionary's interfaces for system access and proof-checking?}
\vspace{\baselineskip}

Untrusted code executing under Supervisionary's supervision could be challenged to prove some theorem each time it wished to open a file on the filesystem, or otherwise perform some side-effect.
These theorems could be correctness or security-related theorems, corresponding to a prevailing \deffont{policy} in force.
Moreover, as Supervisionary is capable of capturing the runtime state of untrusted code, these challenges can be HOL predicates that are functions of the reified runtime state of untrusted code, the kernel, and the arguments, and name of, the invoked system call.

Two predicates of interest are $\lam{k}\lam{u}\lam{s}\top$ and $\lam{k}\lam{u}\lam{s}\bot$.
Here, $\top$ and $\bot$ are truth and falsity, and $k$, $u$, and $s$ the kernel and untrusted code states, and packed system call metadata, reified as HOL data, at the point of invocation of the system call.
One can always prove $\{\} \vdash \top$ and therefore $\lam{k}\lam{u}\lam{s}\top$ represents no restriction, whereas $\{ \} \vdash \bot$ is never provable, absent axioms, with $\lam{k}\lam{u}\lam{s}\bot$ representing a ``closing off'' of a system call.
By making the challenge a function of the system call name and arguments, this closing off can be specific, banning a process from calling a particular system call, or a system call with a particular set of arguments, allowing us to mimic mechanisms like \texttt{seccomp} from Linux.

We can go further: metadata about the behaviour of a running process could be maintained---for example, a record of the system calls invoked by a process, thus far.
This record could be used in forming security or correctness challenges, for example by forcing untrusted code to prove that writes to a socket only ever happen after a read, or reads and writes on sockets satisfy some protocol.
In short, HOL acts as \emph{lingua franca} between kernel and untrusted code, through which arbitrarily complex policies may be expressed.

\deffont{Jailing}, wherein a process voluntarily sheds capabilities, is common in existing operating systems.
This can be captured in Supervisionary by allowing a process to dynamically replace the prevailing policy, $\phi$, with a new policy $\psi$, after proving that $\psi$ is a refinement of $\phi$: $\{\} \vdash \fall{k}\fall{u}\fall{s} \psi\ k\ u\ s \longrightarrow \phi\ k\ u\ s$.
Note that this expresses that the states described by $\psi$ are a subset of those described by $\phi$.

Note that this idea shares some similarities with \deffont{proof carrying code}, wherein binaries are accompanied with (skeleton) proofs of their adherence to some policy, and these proofs checked by the operating system prior to execution~\cite{DBLP:conf/popl/Necula97}.
However, the ideas sketched above generalise this: proofs can be generated dynamically, as the program executes, and could more aptly be called \deffont{proof generating code}.
Processes and the Supervisionary kernel work together to come to a ``mutual understanding'' that the behaviour of the process is indeed concordant with the prevailing policy. 

\paragraph{Restructuring proof-checking tools}

Existing tools in the HOL family could be refactored around Supervisionary by changing their kernels to act as frontends to the Supervisionary kernel for untrusted automation routines.
Arguably, this increases the robustness of existing tools, enforcing separation by isolation, rather than module boundaries.
More interestingly, this also means that Supervisionary acts as a mechanism for ``transporting'' definitions and theorems between systems within the wider HOL family: systems are capable of referring to, and manipulating, kernel objects produced by other systems through Supervisionary kernel handles.
With this, we can also bootstrap a full theorem-proving environment, with associated libraries of content, on top of Supervisionary without writing it from scratch.

Our prototype acts as both programmable proof-checker and general-purpose sandbox.
The previous section highlighted that this idea can be used to enforce runtime properties of programs.
But, we can also use this blurring to import ideas from the operating systems community into the design of proof-checking software itself.
For example, libraries of mathematical theorems and definitions could be presented to Supervisionary's users via a hierarchical or tag-based file-system, and explored with command-line tools in an interactive shell atop the Supervisionary kernel.

\paragraph{Lastly}

Supervisionary's dual interpretation as sandbox and proof-checker blurs the boundary between static and runtime verification, and between proof-checker and sandbox.
Whilst proofs can be generated by executables themselves at runtime, they could also be generated interactively by users, prior to the execution of a program, or even generated for use by a program by other programs.

For example, the operational semantics and instruction decoding functionality of Wasm is clearly embeddable in HOL~\cite{DBLP:conf/cpp/Watt18}.
Using this, properties of a program $P$ to be executed under Supervisionary could be established statically, and registered with the kernel, perhaps interactively by a user or by another program executing before $P$ executes.
This theorem can then be used by $P$ in closing challenges from Supervisionary, issued at runtime, tying the correctness of $P$ to its execution.

\newpage


\end{document}